\documentclass[journal]{IEEEtran}
   \usepackage{graphicx}
\usepackage[cmex10]{amsmath}
\begin{document}
%
\title{Magnetic Generation due to Mass Difference between Charge Carriers}
%
%
%

\author{Shi~Chen,
        Jia~Kun~Dan,
        Zi~Yu~Chen,
        and~Jian~Feng~Li
\thanks{S. Chen is with Institute of Fluid Physics, China Academy of Engineering Physics, Mianyang,
Sichuan, 621900 China e-mail: agog992000@163.com.}
\thanks{J. K. Dan, Z. Y. Chen and J. F. Li are with Institute of Fluid Physics, China Academy of Engineering Physics.}
\thanks{Manuscript received April 19, 2005; revised January 11, 2007.}}

%
%

\markboth{Journal of \LaTeX\ Class Files,~Vol.~6, No.~1, January~2007}%
{Shell \MakeLowercase{\textit{et al.}}: Bare Demo of IEEEtran.cls for Journals}
%



\maketitle

\begin{abstract}
The possibility of spontaneous magnetization due to the ``asymmetry in mass'' of charge carriers in a system is investigated. Analysis shows that when the masses of positive and negative charge carriers are identical, no magnetization is predicted. However, if the masses of two species are different, spontaneous magnetic field would appear, either due to the equipartition of magnetic energy or due to fluctuations together with a feedback mechanism. The conditions for magnetization to occur are also obtained, in the form of $n$-$T$ phase diagram. The theory proposed here, if confirmed by future observations and/or experiments, would provide a new insight on the origin of magnetic fields in the universe.
\end{abstract}


%
\IEEEpeerreviewmaketitle

\section{Introduction}
\label{intro}
%
%
%
%
\IEEEPARstart{T}{he} origin of magnetic fields in the universe has been a long-time question that provokes constant interests of human beings. Magnetic fields exist in almost all astrophysical entities and systems, from main sequence stars like our sun\cite{RefJ:Hale_1908}, to galaxies\cite{RefJ:Kulsrud_2008,RefJ:Reid_1990}, and to even larger scales\cite{RefJ:Heiles_1996}. With developing techniques, such as Faraday rotation\cite{RefJ:Rand_1994,RefJ:Han_2001} and Zeemen-effect\cite{RefJ:Hale_1908,RefJ:Reid_1990,RefJ:Widrow_2002,RefJ:Han_2007} methods, properties of cosmic magnetic fields are being studied more accurately and thoroughly. Progress has also been made on the question of how these magnetic fields are generated. The famous dynamo theory\cite{RefJ:Han_2001,RefJ:Widrow_2002,RefJ:Parker_1955,RefJ:Jiang_2005,RefJ:Bonanno_2006,RefJ:Urpin_2004,RefJ:Bonanno_2003} has provided a mechanism to sustain and amplify some initial ``seed'' magnetic fields. As for the generation of these ``seed'' fields, various models have been proposed (\textit{e.g.} Biermann-battery effect\cite{RefJ:Biermann_1951}, phase-transition in early universe\cite{RefJ:Hogan_1983,RefJ:Turner_1988}, \textit{etc}.), but no common agreements have been achieved. Therefore, the problem of the generation of cosmic magnetic fields is not solved yet.

The universality of existence of magnetic fields in diverse systems implies that its origin might be related to some fundamental principals of physics, especially the one of conservation laws and symmetries\cite{RefB:Goldstein}. Occurrence of a detectable quantity is usually caused by the breakdown of the corresponding symmetry. From this point of view, we notice that a common feature of celestial systems mentioned above is that, the masses of positive and negative charge carriers of which they are made, are different. This ``asymmetry in mass'' could cause difference in motions of two species, and finally result in the generation of magnetic field. In this work, we present analysis following this logic, and point out the possibility of magnetic generation due to the asymmetry in mass of charge carriers.
\section{Theoretical Model}
\label{sec:theory}
The relation between conservation laws and symmetries of a system plays an important, if not essential, role in physics, as stated in Noether's theorem\cite{RefB:Goldstein}. Each kind of symmetry corresponds to a conserved quantity, and vice verse, each symmetry breakdown would lead to a detectable quantity. One famous example is the non-zero chirality due to breakdown of parity symmetry. Likewise, the occurrence of electromagnetic properties of a system could be regarded as the consequence of breakdown of charge symmetry. In this work, we will show that in a system consisting of opposite charge carriers, the asymmetry in mass of positive and negative charge carriers could, under certain conditions, result in spontaneous magnetization.

Consider a system which consists of particles with opposite charge, \textit{i.e.} $+q$ and $-q$, and denote their rest masses by $m_+$ and $m_-$, respectively. Assume all particles have the same temperature $T_0$, and their typical thermal velocities are
\begin{eqnarray}
    \label{equ:1} u_+&=&\sqrt{\frac{k_B T_0}{m_+}},\\
    \label{equ:2} u_-&=&\sqrt{\frac{k_B T_0}{m_-}},
\end{eqnarray}
where $k_B$ is the Boltzmann constant. In thermal equilibrium, and without bulk motions, the averaged velocity of each type of particles is equal to zero, \textit{i.e.}
\begin{eqnarray}
    \label{equ:3} \langle\mathbf{v}_+\rangle&=&0,\\
    \label{equ:4} \langle\mathbf{v}_-\rangle&=&0,
\end{eqnarray}
where $\langle\cdots\rangle$ denotes average over the distribution function. However, the average of the square of velocity, which is proportional to the kinetic energy, is non-zero:
\begin{eqnarray}
    \label{equ:5} \langle\mathbf{v}^2_+\rangle&=&u^2_+ =\frac{k_B T_0}{m_+},\\
    \label{equ:6} \langle\mathbf{v}^2_-\rangle&=&u^2_- =\frac{k_B T_0}{m_-}.
\end{eqnarray}

Since particles are charged, they carry electromagnetic fields with them. Generally, electric fields are easily shielded out by freely-moving charge carriers, while magnetic fields are not so easy to be canceled, thus could exist for a much longer time and in a much wider scope\cite{RefB:Parker}. Therefore, in the rest of discussion, we would focus on magnetic fields in the system and neglect electric fields for simplicity.

	The magnitude of magnetic field generated by a moving charge carrier is proportional to its velocity, \textit{i.e.}
\begin{equation}
    \label{equ:7} B=av,
\end{equation}
where $a$ is some coefficient and is not important in our discussion. Since the directions of velocities of particles are random, so are the directions of magnetic fields and they cancel out:
\begin{eqnarray}
    \label{equ:8} \langle\mathbf{B}_+\rangle&=&0,\\
    \label{equ:9} \langle\mathbf{B}_-\rangle&=&0.
\end{eqnarray}
However, the average of magnetic field energy is non-zero, and is proportional to $v^2$:
\begin{eqnarray}
    \label{equ:10} \langle\mathbf{B}^2_+\rangle&=&b \langle\mathbf{v}^2_+\rangle =b \frac{k_B T_0}{m_+},\\
    \label{equ:11} \langle\mathbf{B}^2_-\rangle&=&b \langle\mathbf{v}^2_-\rangle =b \frac{k_B T_0}{m_-},
\end{eqnarray}
where $b$ is another unimportant coefficient. The critical result here is that the magnetic field energy of one kind of charge carriers is inversely proportional to its mass. If the masses of positive and negative charge carriers are equal, which we call ``symmetry in mass'', then the magnetic field energy of each kind has the same magnitude. The equipartition of magnetic energy is fulfilled, and the system will remain in this equilibrium state, without occurrence of non-zero averaged magnetic field. On the other hand, if the mass of positive charge carriers is not equal to that of negative ones, which we call ``asymmetry in mass'', the magnetic energies of two kinds are not identical. The system will evolve to reduce the difference in the magnetic energy of two species. As a result, spontaneous magnetization occurs. This process will be investigated in the following.

	Denote the number of positive charge carriers by $N$, which is the same as that of negative ones. Initially, the total magnetic energy of the system is:
\begin{equation}
    \label{equ:12} U_B=Nb\frac{k_B T_0}{m_+}+Nb \frac{k_B T_0}{m_-}.
\end{equation}
For simplicity, we neglect the energy exchange between thermal motion and magnetic field, so the total magnetic energy of the system is conserved. Assume in the final state, a portion $c_+$ of positive charge carriers generate non-random, coherent magnetic fields. The field then consists of $(1-c_+)N$ random $\mathbf{B}_{r+}$ and $c_+N$ non-random $\mathbf{B}_{nr+}$. The total field due to positive charge carriers is:
\begin{equation}
    \label{equ:13} B_{\text{total}+}=c_+Na\sqrt{\frac{k_B T_0}{m_+}}.
\end{equation}
Therefore, the magnetic energy of positive particles is:
\begin{equation}
    \label{equ:14} U_{B+}=(1-c_+)Nb\frac{k_B T_0}{m_+} +c^2_+N^2b\frac{k_BT_0}{m_+},
\end{equation}
where the first term of the right-hand-side of Eq.(\ref{equ:14}) is the contribution from the random part of field, while the second term represents the non-random part. Likewise, assume a portion $c_-$ of negative charge carriers create non-random fields. The total field due to negative particles is:
\begin{equation}
    \label{equ:15} B_{\text{total}-}=c_-Na\sqrt{\frac{k_B T_0}{m_-}}.
\end{equation}
The magnetic energy of negative charge carriers is:
\begin{equation}
    \label{equ:16} U_{B-}=(1-c_-)Nb\frac{k_B T_0}{m_-} +c^2_-N^2b\frac{k_BT_0}{m_-},
\end{equation}
Notice that we do not include the term of $\langle \mathbf{B}_{\text{total}+} \cdot \mathbf{B}_{\text{total}-} \rangle$, since the relative orientation between these two vectors is random and the average is zero. The request of equipartition of magnetic energy between positive and negative particles leads to the following equations:
\begin{eqnarray}
    \label{equ:17} (1-c_+)N+c^2_+N^2&=&\frac{N}{2} \left( 1+\frac{m_+}{m_-}\right),\\
    \label{equ:18} (1-c_-)N+c^2_-N^2&=&\frac{N}{2} \left( 1+\frac{m_-}{m_+}\right).
\end{eqnarray}
Denote the ratio of $m_-$ to $m_+$ by $s$, the above equations can be rewritten as:
\begin{eqnarray}
    \label{equ:19} Nc^2_+-c_++\frac{1}{2}-\frac{1}{s}&=&0,\\
    \label{equ:20} Nc^2_--c_-+\frac{1}{2}-s&=&0.
\end{eqnarray}
The existence of real solutions to these equations leads to the following requirement:
\begin{equation}
    \label{equ:21} 1/2<s<2.
\end{equation}
This result is understandable. If the difference in mass is too large, so will be the difference in magnetic energy. Therefore no matter how strong the spontaneous magnetic field is, the equipartition condition could never be fulfilled. If the ratio of positive and negative charge carriers falls in the above range, the portion of $c_i$ ($i=+,-$) is approximately proportional to $N^{-1/2}$, leading to a total magnetization proportional to $N^{1/2}$.

As mentioned above, for the case in which $s>2$ or $s<1/2$, the equipartition of magnetic energy could not be achieved. Under appropriate conditions, the lack of equipartition could lead to instability in the system. Therefore, the simple analysis above is not enough to describe the evolution of the system. A dynamic description is needed, which will be discussed in detail below.

	Without loss of generality, consider the case of $s\ll1/2$, which is easily satisfied in realistic plasmas. The positive charge carriers are much heavier than the negative ones, and their magnetic energy can be neglected compared to that of negative ones. The large difference in mass also means that energy exchange between the two species is relatively inefficient. Therefore, we can approximate the positive particles as a uniform background, and focus on the evolution of negative ones.

	Assume that initially the negative charge carriers are in thermal equilibrium. This state could be unstable due to fluctuations and some microscopic feedback mechanism. The growth of such instability would finally result in magnetization. Fluctuations mainly come from charge carriers with energy higher than $k_BT_0$. These fast carriers penetrate longer distances in the system, and create stronger toroidal magnetic fields around them. If these fields are strong enough, they could trap some of ambient carriers and force them to move along the lines of force, forming local current loops. This magnetic-trapping process transfers part of thermal energy into magnetic energy in the loops, thus leads to a decrease of local temperature. Therefore, fluctuations from fast carriers cause perturbations in thermal and magnetic energy.

	For the instability to develop, perturbations have to be enhanced spontaneously, which requires some feedback process. If the strength of loops is not large enough, or the distance between two neighboring loops is not small enough, local current loops could not interact with each other efficiently. As a result, they will die out due to collisions and their magnetic energy will be transferred into thermal energy again. In this case, perturbations could not develop and the system is stable. On the other hand, if local current loops are both strong enough and close enough to each other, their interaction will increase local magnetic fields. In return, stronger magnetic fields will trap more ambient particles and enhance local currents. In this case, a feedback mechanism exists, due to interactions between current loops. This feedback process enlarges initial perturbations and would lead to instability.

	In order to describe the above-mentioned process, it is convenient to adapt cylindrical coordinates. For simplicity, we consider only one-dimensional case. Quantities depend only on $z$ and $t$. A fast carrier moves along the $z$ axis, generating toroidal magnetic field $B_\theta\mathbf{e}_\theta$, where $\mathbf{e}_\theta$ is the unit vector of angular direction. According to the drift theory\cite{RefB:Chen}, trapped ambient particles move along the line of force, forming a local current $j_\theta\mathbf{e}_\theta$. The generation of local current can be described effectively as:
\begin{equation}
    \label{equ:22} \varepsilon_\theta(z,t)- L\frac{\partial}{\partial t}j_\theta(z,t)= Rj_\theta(z,t),
\end{equation}
where $\varepsilon_\theta$ is effective electromotive force due to $B_\theta$, $L$ and $R$ are effective induction and resistance of the system. $L$ and $R$ have the following approximate forms:
\begin{eqnarray}
    \label{equ:23} L&=&\frac{m_-}{n_-q^2\lambda_D},\\
    \label{equ:24} R&=&\frac{m_-}{n_-q^2\lambda_D\tau_-},
\end{eqnarray}
where $n_-$ is the number density of negative charge carriers, $\tau_-$ is the mean-free-time, and $\lambda_D$ is Debye length. Scaling with the maximum magnitude of $\varepsilon_\theta$ leads to the following equation:
\begin{equation}
    \label{equ:25} \tau_-\frac{\partial}{\partial t} \tilde{j}_\theta(z,t)+\tilde{j}_\theta(z,t)= \tilde{\varepsilon}_\theta(z,t).
\end{equation}
Quantities with tildes are scaled. If we choose $\tau_-$ and $\lambda_D$ as characteristic temporal and spatial scales, Eq.(\ref{equ:25}) can be further simplified as:
\begin{equation}
    \label{equ:26} \frac{\partial}{\partial \tilde{t}} \tilde{j}_\theta(\tilde{z},\tilde{t})+ \tilde{j}_\theta (\tilde{z},\tilde{t})=\tilde{\varepsilon}_\theta (\tilde{z},\tilde{t}).
\end{equation}
Conversion of magnetic energy to thermal energy due to collision can be described by a diffusion term $D\partial^2 \tilde{j}_\theta/\partial \tilde{z}^2$ added to the right-hand-side of Eq.(\ref{equ:26}). The feedback process can be approximated as:
\begin{equation}
    \label{equ:27} \frac{\partial}{\partial \tilde{t}} \tilde{j}_\theta(\tilde{z},\tilde{t})= \eta\tilde{j}_\theta (\tilde{z},\tilde{t}),
\end{equation}
where $\eta$ is the growth rate and depends on the magnitude of local current $\tilde{j}_\theta$. Therefore, the complete equation has the following form:
\begin{equation}
    \label{equ:28} \frac{\partial}{\partial \tilde{t}} \tilde{j}_\theta(\tilde{z},\tilde{t})+ \tilde{j}_\theta (\tilde{z},\tilde{t})=\tilde{\varepsilon}_\theta (\tilde{z},\tilde{t})+ D\frac{\partial^2 \tilde{j}_\theta(\tilde{z},\tilde{t})}{\partial \tilde{z}^2} +\eta\tilde{j}_\theta(\tilde{z},\tilde{t}).
\end{equation}

Solving Eq.(\ref{equ:28}) shows that both stable and unstable solutions exist, depending on the competition between diffusion and feedback processes. Detailed numerical results will be presented in the next section.
\section{Numerical Results}
\label{sec:numresults}
The equation (\ref{equ:28}) derived in Sec.\ref{sec:theory} contains the effects of fluctuation, diffusion, as well as feedback. In this section, we will discuss each of these effects and present detailed results. For convenience, all tildes will be omitted.
\subsection{Fluctuation}
\label{sec:NR.1}
To study the effect of fluctuations, we can set the diffusion and feedback coefficients to zero. The resulting equation is just Eq.(\ref{equ:26}). The source of fluctuations is a fast charge carrier, which can be modeled by:
\begin{equation}
    \label{equ:29} \varepsilon_\theta(z,t)= \exp{\left(-\frac{(z-z_0-v_0t)^2}{\Delta^2_z}\right)} \exp{\left(-\frac{(t-t_0)^2}{\Delta^2_t}\right)},
\end{equation}
where $z_0$ and $t_0$ are initial position and time of the fast charge carrier, $v_0$ is its velocity, $\Delta_z$ and $\Delta_t$ are spatial width and temporal duration of the charge carrier. This assumption indicates that the charge is a Gaussian wave-package in space, and also is a Gaussian pulse in duration.

\begin{figure}
  \includegraphics[width=\columnwidth]{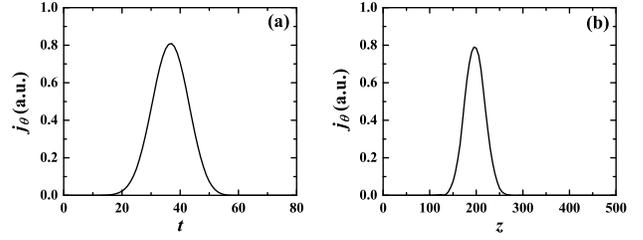}
  \caption{Typical solution of Eq.(\ref{equ:26}). The total length and time of calculations are $z_{\text{max}}=500$, and $t_{\text{max}}=500$. The velocity of the charge carrier is $v_0=3$. The spatial width $\Delta_z=30$, and the duration $\Delta_t=0.1z_{\text{max}}/v_0$. The initial position and time are $z_0=0.1z_{\text{max}}$, and $t_0=0.25z_{\text{max}}/v_0$. The current at a certain position (\textit{e.g.} at $z=0.3z_{\text{max}}$, as shown in (a)) evolves in a Gaussian-like way. The spatial dependence of current (\textit{e.g.} at $t=0.1t_{\text{max}}$, as shown in (b)) is also Gaussian.}\label{fig:fluc}
\end{figure}

The current fluctuation caused by such a fast charge carrier, as shown in Fig.\ref{fig:fluc}, is pulsed and local. The temporal and spatial behaviors are both Gaussian-like, indicating that the fluctuation is not only short-living within the duration of the source, but also localized in the neighborhood of the charge carrier. Without diffusion, the fluctuation will not spread out in space. Without feedback, it will not grow spontaneously. These results are in consistent with our knowledge of thermal fluctuations.
\subsection{Feedback}
\label{sec:NR.2}
\begin{figure}
  \includegraphics[width=\columnwidth]{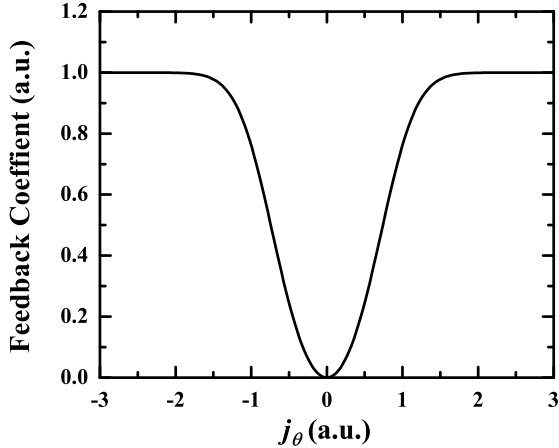}
  \caption{Feedback coefficient, scaled by $\eta_0$. When the local current is zero, no feedback process would occur, which means the coefficient is zero. When local current loops are totally aligned, the coefficient reaches a maximum value.}
\label{fig:feedback-co}
\end{figure}

Feedback process is essential for spontaneous magnetization. It stems from the interaction between local current loops. When there is no net magnetic field, the directions of current loops are random, and the average field due to them is zero. However, neighboring loops have the tendency of aligning in the same direction, through magnetic interaction between them. Such alignment of loops generates non-zero average field and this field strengthens the alignment. Therefore, the strength of this feedback mechanism is zero if all loops are randomly directed, and reaches a maximum value when all loops are aligned in exactly the same direction. Thus, we can approximate the coefficient of the feedback by the following relation:
\begin{equation}
    \label{equ:30} \eta(j_\theta)=\eta_0\left(2\frac{\exp{(j^2_\theta)}}{\exp{(j^2_\theta)} +\exp{(-j^2_\theta)}} -1\right),
\end{equation}
where $\eta_0$ is the characteristic strength of the feedback. The dependence of this coefficient on current is shown in Fig.\ref{fig:feedback-co}.

\begin{figure}
  \includegraphics[width=\columnwidth]{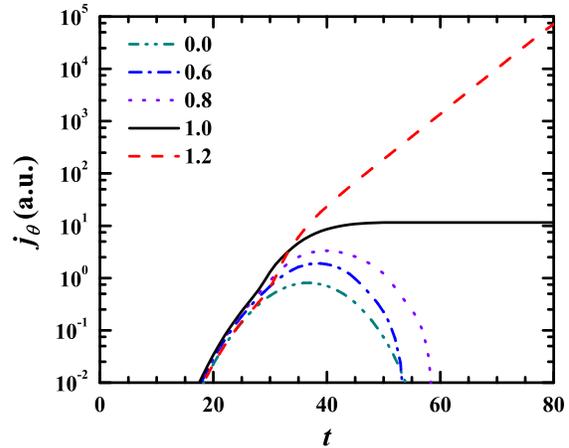}
  \caption{Current dependence on time at $z=0.3z_{\text{max}}$. The calculation parameters are the same as those used in Fig.\ref{fig:fluc}, except for $\eta_0$. The current is in $\log$ coordinate. Results corresponding to feedback coefficient of 0.0(dash-dot-doted), 0.6(dash-doted), 0.8(doted), 1.0(solid), and 1.2(dashed) are presented.}
\label{fig:feedback}
\end{figure}

With this approximation, we turn on the feedback, while keep the diffusion term turned off, and calculate local current under the same parameters as above. The results, as shown in Fig.\ref{fig:feedback}, imply that there are several effects caused by feedback.

First, feedback process increases the maximum current that can be generated by fluctuations, which can be seen obviously from lines with feedback coefficient of 0.6(dash-doted), 0.8(doted), and 1.0(solid) in Fig.\ref{fig:feedback}. Secondly, and more importantly, the duration of fluctuated current is enhanced, especially for large feedback coefficients. For $\eta_0=0$ and $\eta_0=0.6$ (dash-dot-doted and dash-doted lines in Fig.\ref{fig:feedback}), the durations are both roughly 33. When $\eta_0$ increases to 0.8 (doted line in Fig.\ref{fig:feedback}), the duration expands to about 40. When $\eta_0=1.0$ (solid line in Fig.\ref{fig:feedback}), which corresponds to certain threshold value, the duration is longer than the total calculation time (in this case, 500). Thirdly, as $\eta_0$ increases to a value larger than the threshold (\textit{e.g.} $\eta_0=1.2$), fluctuated current will grow exponentially (dashed curve in Fig.\ref{fig:feedback}), even after the initial perturbing source has ceased. In other words, spontaneous magnetization occurs.
\subsection{Diffusion}
\label{sec:NR.3}
\begin{figure}
  \includegraphics[width=\columnwidth]{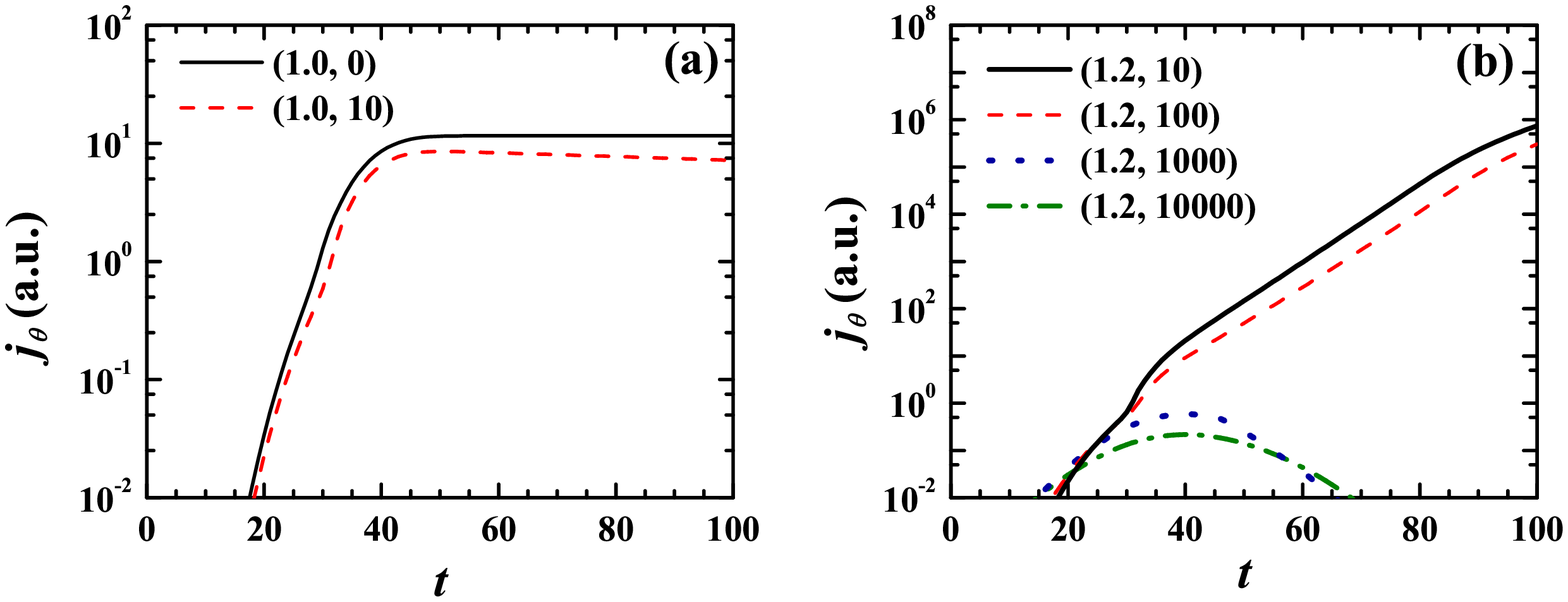}
  \caption{Diffusion effect on fluctuated current at $z=0.3z_{\text{max}}$. The calculation parameters are identical to those used above. The feedback coefficient is 1.0 in (a) and 1.2 in (b). If $D$ is small, \textit{e.g.} $D=0$(solid) and $D=10$(dashed) in (a), and $D=10$(solid) and $D=100$(dashed) in (b), only minor corrections occur. However, for large $D$, as $D=1000$(doted) and $D=10000$(dash-doted) in (b), exponential growths disappear.}
\label{fig:feeddiff}
\end{figure}

The effect of diffusion depends on its coefficient. At small diffusion coefficients, minor corrections of fluctuated current (see Fig.\ref{fig:feeddiff}) occur. However, if the diffusion coefficient is large, it can destroy the exponential growth of current (see Fig.\ref{fig:feeddiff}(b)), and as a result, stop spontaneous magnetization.

%
\section{Phase Diagram}
\label{sec:phd}
Previous analysis indicates that feedback process, which is essentially the formation of and interaction between local current loops, is critical for the occurrence of spontaneous magnetization. In this section, we will discuss the conditions under which the above-mentioned mechanism works. When expressed as relations between the temperature and number density of charge carriers in the system, these conditions mark the spontaneous magnetization area in the $n$-$T$ phase diagram.
\subsection{High Temperature and Density Regime}
\label{sec:phd.1}
In high temperature and density regime, the formation of local current loops depends on magnetic trapping of ambient charge carriers by toroidal magnetic field generated by a fast one. To the lowest order, gradient and curvature drift theory gives good approximations of particle trajectories, which confirm the existence of current loops. The requirement that ambient carriers could be trapped, which is also the requirement that drift theory is valid, is that the thermal energy of ambient carriers is less than their magnetic energy in the field. Assuming the relativistic factors of the fast charge carrier are $\beta$ and $\gamma$, then this requirement can be written as:
\begin{equation}
    \label{equ:31} k_BT<\frac{q^4\gamma^2\beta^2}{m_-c^2\varepsilon^2_0} n^{2/3}_-,
\end{equation}
where $c$ is the speed of light in vacuum and $\varepsilon_0$ is the vacuum permittivity. Since the velocity of the fast carrier is close to $c$, $\beta$ is nearly 1. If we suppose the energy of the fast carrier is ten times that of thermal energy, the above relation reduces to:
\begin{equation}
    \label{equ:32} k_BT>\frac{(m_-c^2)^3 \varepsilon^2_0}{100q^4}n^{-2/3}_-.
\end{equation}
This condition will be referred to as ``\textit{magnetic trapping}'' condition.

Secondly, the characteristic strength of feedback, $\eta_0$, is related to the growth time of local current. In general, if the growth time is short comparing to the mean-free-time $\tau_-$, feedback could establish adequately and $\eta_0$ is large. On the other hand, if the growth time is longer than $\tau_-$, collisions would interrupt the interaction process and $\eta_0$ remains small. Therefore, we approximate that $\eta_0$ is inversely proportional to the growth time, which is scaled with $\tau_-$. Causality requires that the characteristic growth time of local current should be longer than $\lambda_D/c$. Take this time as an estimate of the growth time, and notice that from previous calculations $\eta_0>1$ for exponential growth to occur. This leads to another relation:
\begin{equation}
    \label{equ:33} \frac{c\tau_-}{\lambda_D}>1.
\end{equation}
Substituting standard statistical physics model for $\tau_-$ and plasma physics model for $\lambda_D$, this relation has the following form:
\begin{equation}
    \label{equ:34} k_BT>\sqrt{\frac{q^6}{\varepsilon^3_0 (m_-c^2)}}n^{1/2}_-,
\end{equation}
and will be referred to as ``\textit{feedback coefficient}'' condition.

Furthermore, in the whole analysis we neglect thermal radiation loss of the system. When the typical frequency of thermal radiation is smaller than the plasma frequency of the system, radiation loss occurs only on the boundary and can be ignored. However, if the typical frequency of thermal radiation is greater than the plasma frequency, the system is transparent to thermal radiation. In this case, radiation loss cannot be dropped. Therefore, the condition under which we can neglect radiation effect is:
\begin{equation}
    \label{equ:35} k_BT<\hbar \sqrt{\frac{q^2}{\varepsilon_0m_-}}n^{1/2}_-,
\end{equation}
where $\hbar$ is the Plank constant. This condition will be called ``\textit{thermal radiation}'' condition. Notice that up to now we have used the rest mass $m_-$ of the negative charge carrier, which means we have neglected relativistic effect of ambient particles. If the temperature is extremely high, ambient particles would also be relativistic. Therefore Eqs. (\ref{equ:31}) to (\ref{equ:35}) need to be modified.

\begin{figure}
  \includegraphics[width=\columnwidth]{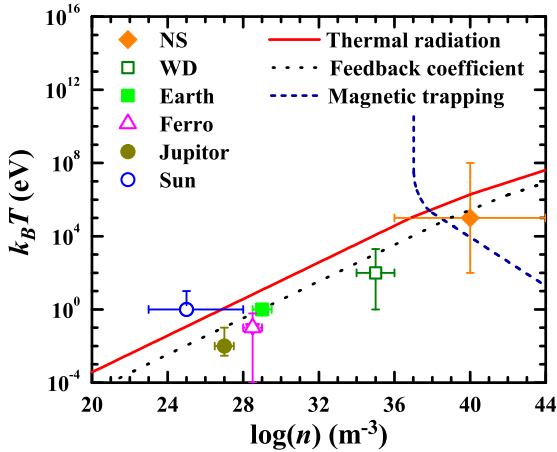}
  \caption{Phase diagram for spontaneous magnetization in both high and low temperature/density regimes. The negative charge carriers in the system are electrons. In the high temperature/density regime, spontaneous magnetization happens in the area above magnetic trapping (dashed) and feedback coefficient condition (doted) lines, and below thermal radiation condition (solid) line. Typical density and temperature of neutron stars (NS, diamond) and white dwarfs (WD, open square) are located in or near the predicted magnetization area. In the low temperature/density case, only feedback coefficient (doted) and thermal radiation (solid) conditions are included. The temperature/density parameters of the Sun (open circle), the Earth (solid square), typical ferromagnetic materials (open triangle), and Jupiter (solid circle) are also shown.}
\label{fig:phase}
\end{figure}

Combination of magnetic trapping, feedback coefficient, and thermal radiation conditions gives out the parametric range for spontaneous magnetization in $n$-$T$ phase diagram. Taking electrons as the negative charge carriers in the system, we obtain the phase diagram shown in Fig.\ref{fig:phase}. Notice that for the above process to occur, there is a minimum number density of about $10^{38}$m$^{-3}$. This lower bound of density is much higher than what we can achieve in laboratories, but is not unacceptable in astrophysical context. Compact celestial bodies, such as white dwarfs (open square in Fig.\ref{fig:phase}) and neutron stars (diamond in Fig.\ref{fig:phase}), have densities comparable to, and even higher than this value, not to mention the early epoch of universe. Typical transition temperature is in the range from tens of keV to MeV. The exact values of transition temperature are unimportant here, because our analysis is much simplified and idealized. The important thing is that spontaneous magnetization could happen not only at a low temperature (\textit{e.g.} in ferromagnetic materials), but also at a very high temperature. Applying our model to calculate magnetic fields of neutron stars, which have particle number densities ranging from $10^{36}$ to $10^{44}$ m$^{-3}$, we obtain magnetic field strength ranging from $10^{10}$ to $10^{15}$ G. This range covers the most observed magnetic fields of known neutron stars and magnetars\cite{RefB:Lyne,RefJ:Thompson_1993}.

\subsection{Low Temperature and Density Regime}
\label{sec:phd.2}
In the range of low temperature and density, the mechanism mentioned above still works, as long as some minor corrections are considered. In high temperature/density case, the magnetic field generated by fast-moving particles is responsible for the formation of local current loops. On the other hand, in low temperature/density case, atomic magnetic moments replace the role of local current. Therefore, fast-moving particles just introduce fluctuations, and the ``magnetic trapping'' condition is not necessary anymore. The resulting phase diagram is shown in Fig.\ref{fig:phase}, together with the temperature/density parameters of the Sun (open circle), the Earth (solid square), typical ferromagnetic materials (open triangle), and Jupiter (solid circle). We can see again that our simple analysis agrees well with the knowledge of magnetism in this range of parameters.
\subsection{Intermediate Temperature and Density Regime}
\label{sec:phd.3}
The high and low temperature/density regimes are two extremes, and thus can be treated by our idealized theory. In the intermediate range, however, various physical processes occur and interact with one another. Therefore, it is no longer a good approximation to neglect any one of the processes or effects. Simple analysis is not enough and more sophisticated and delicate theories are needed.

Despite the inadequateness of an ideal model, we could still find clues to how the phase diagram might look like in this regime. No matter how complicate the situation might be, the competition between random thermal motion of and interaction between particles is essential. In order for the magnetization to occur, feedback mechanism has to overcome thermal motions, which means for a given density, there would be a maximum phase-transition temperature. From previous analysis and calculations, it is likely that the areas of magnetization in high, intermediate, and low temperature/density regimes are connected, and form an integrated area in the phase diagram. The identification of the boundary of this whole area will be a future problem.
\section{Discussions}
\label{sec:disc}
\subsection{Magnetic Generation in Early Universe}
\label{sec:D.1}
One application of our model is to estimate magnetic generation in the early epoch of universe. Assume a one-cubic-centimeter sphere, containing all the known matters in the universe. Assume the particle number density is roughly $10^{87}$m$^{-3}$. As the big bang theory implies, the temperature is also extremely high. Therefore, the condition falls into the above-mentioned high temperature/density regime. Assume that the magnitude of charge of particles is one third of the electron charge. According to Eqs. (\ref{equ:34}) and (\ref{equ:35}) and including relativistic corrections, the temperature range for spontaneous magnetization is roughly from $2.0\times10^{20}$eV to $4.3\times10^{21}$eV. Assuming that magnetization process began at the higher boundary value and then stopped at the lower boundary value, the total magnetic energy gained from this transition is about $4.1\times10^{21}$eV per particle.Assuming the sphere is totally magnetized, and neglecting all the complexities that might occur, this magnetic energy corresponds to a magnetic field of $9.0\times10^{45}$G. If this phase transition occurred before the inflation period, the exponential expansion of the universe due to inflation would reduce the magnitude of magnetic field dramatically. According to the inflation theory, the scale of the universe was enlarged by $10^{26}$ times after inflation. As magnetic flux is conserved, this leads to an overall magnetic field of $0.9\mu$G for present cosmos, which agrees well with observational data of large-scale magnetic fields in the universe\cite{RefJ:Yamazaki_2010}. This estimate is an upper boundary value, due to several reasons. First, since the detailed magnetization process is unknown, the assumption that the sphere is totally magnetized is optimistic. Instead, domains with magnetic fields pointing in different directions might as well appear, resulting in a reduced magnitude of the total magnetic field. Furthermore, during the period of inflation, part of the magnetic energy could convert into the kinetic energy of expansion. Besides, electromagnetic radiation could also take away some magnetic energy. Therefore, all the processes mentioned above, together with some other possible mechanisms, might reduce the nowadays magnetic field strength to less than $0.9\mu$G. The detailed analysis about these processes lay out of the scope of this paper. However, we still consider this agreement as a supportive evidence for our model. Therefore, by assuming that magnetic generation occurred before the inflation instead of after it, we provide an alternative possibility of cosmic magnetic origin\cite{RefJ:Kulsrud_1997,RefJ:Dimopoulos_1998,RefJ:Beltran_2011}.
\subsection{Energy Conversion and Possible Consequences}
\label{sec:D.2}
As mentioned in Sec.\ref{sec:theory}, under certain conditions, spontaneous magnetization process occurs, resulting in an increase in magnetic energy density. Since in our model, only magnetic and thermal energy are considered, energy conservation implies that thermal energy density would decrease accordingly. In the above analysis, conversion between magnetic and thermal energy is neglected, but its effect is readily to be shown. As thermal energy density decreases, the intensity of random thermal motions of charge carriers are reduced, which makes it more difficult to break the ordered local current loops. Therefore, decrease of thermal energy density, in other words, decrease of temperature, is in favor of magnetization, and including energy conversion will not alter the above process qualitatively.

\begin{figure}
  \includegraphics[width=\columnwidth]{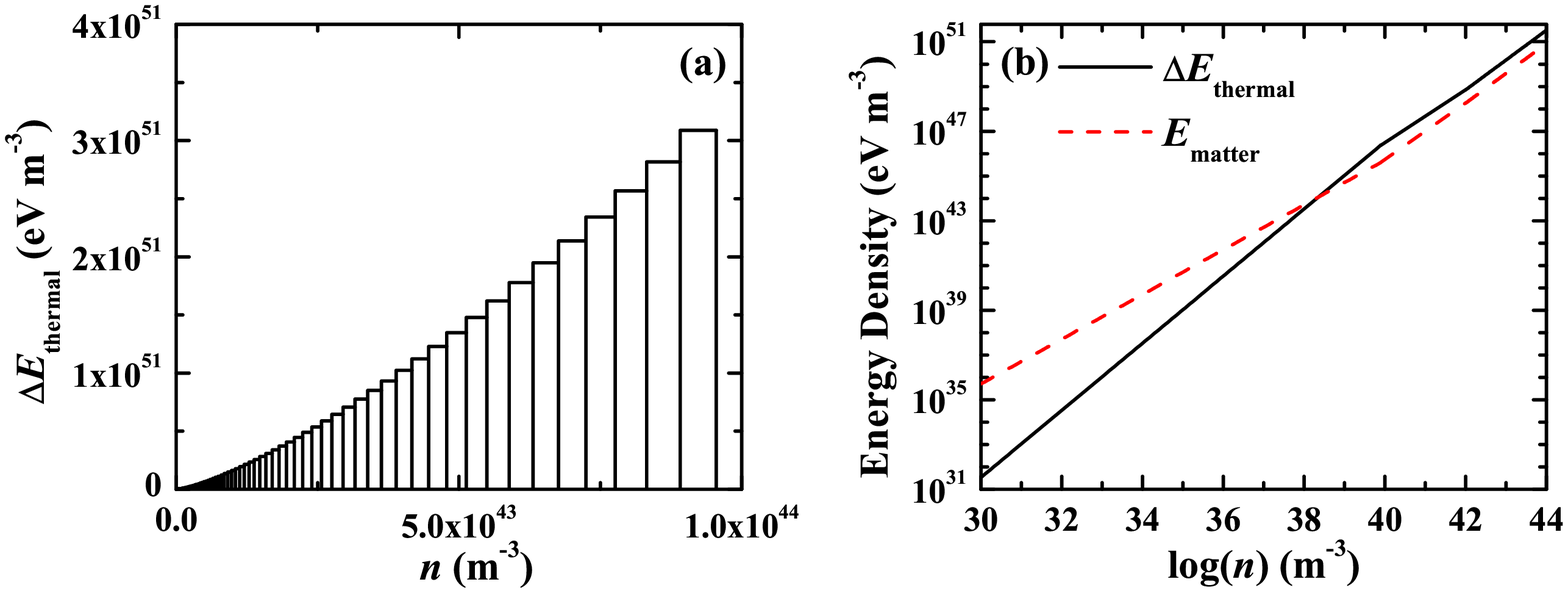}
  \caption{Decrease of thermal energy density. Such decrease($\Delta E_{\text{thermal}}$) grows as $n^{3/2}$, where $n$ is the number density of charge carriers in the system, as shown in (a). In the low temperature/density regime, the amplitude of $\Delta E_{\text{thermal}}$ (solid) is small compared to the mass energy density $E_{\text{matter}}$ (dashed), as shown in (b). However, when $n$ exceeds certain threshold value, $\Delta E_{\text{thermal}}$ could become much larger than $E_{\text{matter}}$.}
\label{fig:energydiff}
\end{figure}

In order to calculate quantitatively the decrease of thermal energy density $\Delta E_{\text{thermal}}$, we assume that the difference between the higher and lower boundary values of temperature at a certain density $n$ in the phase diagram is an approximation of such decrease per one particle. Multiplying the corresponding $n$ will give $\Delta E_{\text{thermal}}(n)$, and the result is shown in Fig.\ref{fig:energydiff}(a). Notice that $\Delta E_{\text{thermal}}$ is proportional to $n^{3/2}$. In the low temperature/density regime, $n$ is relatively small, and $\Delta E_{\text{thermal}}$ is much smaller than the mass energy density, $E_{\text{matter}}$. However, when $n$ exceeds a certain threshold, $\Delta E_{\text{thermal}}$ could be much larger than $E_{\text{matter}}$ (as shown in Fig.\ref{fig:energydiff}(b)). In other words, the magnetic energy density would increase and exceed the total mass energy density at high temperature/density regime. In the early epoch of universe, the temperature and density are very large, indicating that once spontaneous magnetization happens, the magnetic energy would be the overwhelming energy. Since magnetic field exerts out-going pressure through out the system, such large magnetic energy density would cause considerable expansion of the universe. This process might provide some new insight on inflation\cite{RefJ:Guth_1981}.
\section{Conclusion}
\label{sec:con}
In this work, we have investigated the possibility of spontaneous magnetization due to the ``asymmetry in mass'' of charge carriers in a system. When the masses of positive and negative charge carriers are identical, no magnetization is predicted. If the masses of two species are slightly different, equipartition of magnetic energy requires a spontaneous magnetic field, the magnitude of which is proportional to the square root of the total particle number. If the mass difference is large, equipartition of magnetic energy can never be fulfilled. Under appropriate conditions, fluctuations together with a feedback mechanism could result in spontaneous magnetization. The parametric range for this process to occur in the $n$-$T$ phase diagram is also obtained. If this process occurs at an early time in the history of universe, the growth of magnetic energy density could provide an alternative explanation of magnetic generation. Besides, magnetic pressure might play an important role in inflation.

	The main conclusion of this work is the possibility of spontaneous magnetization at a relatively high temperature and density. In order to determine transition conditions more accurately, sophisticated and more realistic models are needed. Since the predicted transition occurs at a temperature and density range beyond the reach of nowadays experimental ability, the verification of the theory relies on astrophysical data, especially those about cosmic magnetic fields and properties of compact celestial bodies. The theory proposed in this article, if confirmed by future observations and/or experiments, would provide a new sight on the origin of magnetic fields in the universe, as well as the essence of inflation.
\section*{Acknowledgment}
\label{sec:ack}
We gratefully acknowledge provoking discussions with all members in the Moonlight Society.
\ifCLASSOPTIONcaptionsoff
  \newpage
\fi


\begin{thebibliography}{1}


%
\bibitem{RefJ:Hale_1908}
G.E. Hale, Astrophys. J. \textbf{28}, (1908) 315
%
\bibitem{RefJ:Kulsrud_2008}
R.M. Kulsrud, E.G. Zweibel, Rep. Prog. Phys. \textbf{71}, (2008) 046901
%
\bibitem{RefJ:Reid_1990}
M.J. Reid, E.M. Silverstein, Astrophys. J. \textbf{361}, (1990) 483
%
\bibitem{RefJ:Heiles_1996}
C. Heiles, Astrophys. J. \textbf{462}, (1996) 316
%
\bibitem{RefJ:Rand_1994}
R.J. Rand, A.G. Lyne, Mon. Not. Roy. Astron. Soc. \textbf{268}, (1994) 497
%
\bibitem{RefJ:Han_2001}
J.L. Han, Prog. Astron. \textbf{19}, (2001) 201
%
\bibitem{RefJ:Widrow_2002}
L.M. Widrow, Rev. Mod. Phys. \textbf{74}, (2002) 775
%
\bibitem{RefJ:Han_2007}
J.L. Han, Chin. J. Nat. \textbf{29}, (2007) 96
%
\bibitem{RefJ:Parker_1955}
E.N. Parker, Astrophys. J. \textbf{122}, (1955) 293
%
\bibitem{RefJ:Jiang_2005}
J. Jiang, J.X. Wang, Prog. Astron. \textbf{23}, (2005) 121
%
\bibitem{RefJ:Bonanno_2006}
A. Bonanno, V. Urpin, G. Belvedere, Astron. Astrophys. \textbf{451}, (2006) 1049
%
\bibitem{RefJ:Urpin_2004}
V. Urpin, J. Gil, Astron. Astrophys. \textbf{415}, (2004) 305
%
\bibitem{RefJ:Bonanno_2003}
A. Bonanno, L. Rezzolla, V. Urpin, Astron. Astrophys. \textbf{410}, (2003) L33
%
\bibitem{RefJ:Biermann_1951}
L. Biermann, A. Schluter, Phys. Rev. \textbf{82}, (1951) 863
%
\bibitem{RefJ:Hogan_1983}
C.J. Hogan, Phys. Rev. Lett. \textbf{51}, (1983) 1488
%
\bibitem{RefJ:Turner_1988}
M.S. Turner, L.M. Widrow, Phys. Rev. D \textbf{37}, (1988) 2743
%
\bibitem{RefB:Goldstein}
H. Goldstein, C. Poole, J. Safko, \textit{Classical Mechanics} (Higher Education Press, Beijing 2005) 344
%
\bibitem{RefB:Parker}
E.N. Parker, \textit{Cosmical Magnetic Fields: Their Origin and Their Activity} (Clarendon/Oxford University Press, Oxford/New York 1979) 22
%
\bibitem{RefB:Chen}
F.F. Chen, \textit{Introduction to Plasma Physics and Controlled Fusion} (Plenum Press, New York 1984) 26
%
\bibitem{RefB:Lyne}
A.G. Lyne, F. Graham-Smith, \textit{Pulsar Astronomy} (Cambridge University Press, Cambridge 2006) 35
%
\bibitem{RefJ:Thompson_1993}
C. Thompson, R.C. Duncan, Astrophys. J. \textbf{408}, (1993) 194
%
\bibitem{RefJ:Dimopoulos_1998}
K. Dimopoulos, Phys. Rev. D \textbf{57}, (1998) 4629
%
\bibitem{RefJ:Yamazaki_2010}
D.G. Yamazaki, K. Ichiki, T. Kajino, \textit{et al.}, Phys. Rev. D \textbf{81}, (2010) 023008
%
\bibitem{RefJ:Kulsrud_1997}
R.M. Kulsrud, R. Cen, J.P. Ostriker, \textit{et al.}, Astrophys. J. \textbf{480}, (1997) 481
%
\bibitem{RefJ:Beltran_2011}
J.J. Beltran, A.L. Maroto, Phys. Rev. D \textbf{83}, (2011) 023514
%
\bibitem{RefJ:Guth_1981}
A. Guth, Phys. Rev. D \textbf{23}, (1981) 347
%
\end{thebibliography}
\end{document}